%% file: apt.tex
\documentclass[12pt]{article}
\include{mydefinitions}

\normalbaselines

\usepackage{natbib}

 \usepackage{sgame}
\usepackage{graphics}

 \newtheorem{theorem}{Theorem}[section]
 \newtheorem{defined}[theorem]{Definition}
 
  \newtheorem{exa}[theorem]{Example}
  \newenvironment{example}{\begin{exa} \rm}{\end{exa}}

 \newtheorem{lemma}[theorem]{Lemma}
 \newtheorem{corollary}[theorem]{Corollary}
 
\newtheorem{note}[theorem]{Note}
\newtheorem{exe}{Exercise}
\newenvironment{exercise}{\begin{exe} \rm }{\end{exe}}

 \author{Krzysztof R. Apt\footnote{Centrum for
     Mathematics and Computer Science (CWI), Science Park 123, 1098 XG
     Amsterdam, the Netherlands,
     and University of Amsterdam.
}}

  \title{A Primer on Strategic Games}

\begin{document}

\date{}



\maketitle

\begin{abstract}
  This is a short introduction to the subject of strategic games.  We
  focus on the concepts of best response, Nash equilibrium, strict and
  weak dominance, and mixed strategies, and study the relation between
  these concepts in the context of the iterated elimination of
  strategies.  Also, we discuss some variants of the original
  definition of a strategic game. Finally, we introduce the basics of
  mechanism design and use pre-Bayesian games to explain it.
\end{abstract}

\section{Introduction}
\label{sec:intro}

Mathematical game theory, as launched by Von Neumann and Morgenstern
in their seminal book, \cite{vNM44}, followed by Nash's contributions
\cite{Nas50,Nas51}, has become a standard tool in economics for
the study and description of various economic processes, including
competition, cooperation, collusion, strategic behaviour and
bargaining.  Since then it has also been successfully used in biology,
political sciences, psychology and sociology. With the advent of the
Internet game theory became increasingly relevant in computer science.

One of the main areas in game theory are \bfe{strategic games}
(sometimes also called \bfe{non-cooperative games}), which form a
simple model of interaction between profit maximising players.  In
strategic games each player has a payoff function that he aims to
maximise and the value of this function depends on the decisions taken
\emph{simultaneously} by all players. Such a simple description is
still amenable to various interpretations, depending on the
assumptions about the existence of \emph{private information}.  The
purpose of this primer is to provide a simple introduction to the most
common concepts used in strategic games: best response, Nash
equilibrium, dominated strategies and mixed strategies and to clarify
the relation between these concepts.

In the first part we consider the case of games with \emph{complete
  information}. In the second part we discuss strategic games with
\emph{incomplete information}, by introducing first the basics of the
theory of \emph{mechanism design} that deals with ways of preventing
\emph{strategic behaviour}, i.e., manipulations aiming at maximising
one's profit.  We focus on the concepts, examples and results, and
leave simple proofs as exercises.

\section{Basic concepts}
\label{sec:basic}

Assume a set $\{1, \ldots, n\}$ of players, where $n > 1$.  A \bfe{strategic game}\index{game!strategic}
 (or
\bfe{non-cooperative game})\index{non-cooperative game|see{strategic game}}
for $n$
players, written as $(S_1, \ldots, S_n, p_1, \ldots, p_n)$, consists of

\begin{itemize}
\item a non-empty (possibly infinite) set $S_i$ of \bfe{strategies},\index{strategy}

\item a \bfei{payoff function} $p_i : S_1 \times \ldots \times S_n \myra \mathbb{R}$,
\end{itemize}
for each player $i$.

We study strategic games under the following basic assumptions:
\begin{itemize}

\item players choose their strategies \emph{simultaneously}; 
subsequently each player receives a payoff from the
resulting joint strategy, 

\item each player is \bfe{rational},\index{rational player}
 which means that his objective is to
maximise his payoff,

\item players have \bfei{common knowledge} of the game and of each
  others' rationality.\footnote{Intuitively, common knowledge of some
    fact means that everybody knows it, everybody knows that
    everybody knows it, etc.}
\end{itemize}

Here are three classic examples of strategic two-player games to
which we shall return in a moment.  We represent such games in the
form of a bimatrix, the entries of which are the corresponding payoffs
to the row and column players.  
\II

\NI
\textbf{Prisoner's Dilemma} 

\begin{center}
\begin{game}{2}{2}
       & $C$    & $D$\\
$C$   &$2,2$   &$0,3$\\
$D$   &$3,0$   &$1,1$
\end{game}
\end{center}

\NI
\textbf{Battle of the Sexes}

\begin{center}
\begin{game}{2}{2}
       & $F$    & $B$\\
$F$   &$2,1$   &$0,0$\\
$B$   &$0,0$   &$1,2$
\end{game}
\end{center}

\NI
\textbf{Matching Pennies} 

\begin{center}
\begin{game}{2}{2}
      & $H$    & $T$\\
$H$   &$\phantom{-}1,-1$   &$-1,\phantom{-}1$\\
$T$   &$-1,\phantom{-}1$   &$\phantom{-}1,-1$
\end{game}
\end{center}
\II

We introduce now some basic notions that will allow us to discuss and
analyse strategic games in a meaningful way.  Fix a strategic game
\[
(S_1, \ldots, S_n, p_1, \ldots, p_n).
\]
We denote $S_1 \times \ldots \times
S_n$ by $S$, call each element $s \in S$ a \bfe{joint strategy},\index{strategy!joint}
or a \bfei{strategy profile},
denote the $i$th element of $s$ by $s_i$, and abbreviate the sequence
$(s_{j})_{j \neq i}$ to $s_{-i}$. Occasionally we write $(s_i,
s_{-i})$ instead of $s$.  Finally, we abbreviate $\times_{j \neq i}
S_j$ to $S_{-i}$ and use the `$_{-i}$' notation for other sequences
and Cartesian products.


We call a strategy $s_i$ of player $i$ a \bfei{best response} to a 
joint strategy $s_{-i}$ of his opponents if
\[
\fa s'_i \in S_i \ p_i(s_i, s_{-i}) \geq p_i(s'_i, s_{-i}).
\]

Next, we call a joint strategy $s$ a \bfei{Nash equilibrium} if each
$s_i$ is a best response to $s_{-i}$, that is, if
\[
\fa i \in \{1, \ldots, n\} \ \fa s'_i \in S_i \ p_i(s_i, s_{-i}) \geq p_i(s'_i, s_{-i}).
\]
So a joint strategy is a Nash equilibrium if no player can
achieve a higher payoff by \emph{unilaterally}
switching to another strategy. 

Finally, we call a joint strategy $s$ \bfe{Pareto
  efficient}\index{Pareto efficient joint strategy} if for no joint
strategy~$s'$
\[
\mbox{$\fa i \in \{1, \ldots, n\} \ p_i(s') \geq p_i(s)$ and $\te i \in \{1, \ldots, n\} \ p_i(s') > p_i(s)$.}
\]
That is, a joint strategy is Pareto efficient if no joint strategy is both
a weakly better outcome for all players and a strictly better outcome
for some player.

Some games, like the Prisoner's Dilemma, have a unique Nash
equilibrium, namely $(D,D)$, while some other ones, like the Matching
Pennies, have no Nash equilibrium. Yet other games, like the Battle of
the Sexes, have multiple Nash equilibria, namely $(F,F)$ and $(B,B)$.
One of the peculiarities of the Prisoner's Dilemma game is that its
Nash equilibrium is the only outcome that is not Pareto efficient.

Let us return now to our analysis of an arbitrary strategic game
$(S_1, \ldots,S_n, p_1, \ldots, p_n)$.
Let $s_i, s'_i$ be strategies of player $i$.
We say that $s_i$ 
 \bfe{strictly dominates $s'_i$}\index{strict dominance}
 (or
equivalently, that $s'_i$ is
 \bfe{strictly dominated by $s_i$}) if
\[
\fa s_{-i} \in S_{-i} \ p_i(s_i, s_{-i}) > p_i(s'_i, s_{-i}),
\]
that $s_i$ \bfe{weakly dominates $s'_i$}\index{weak dominance} (or
equivalently, that $s'_i$ is \bfe{weakly dominated by $s_i$}) if
\[
\mbox{$\fa s_{-i} \in S_{-i} \ p_i(s_i, s_{-i}) \geq p_i(s'_i, s_{-i})$ and
$\te s_{-i} \in S_{-i} \ p_i(s_i, s_{-i}) > p_i(s'_i, s_{-i})$},
\]
and that $s_i$ 
 \bfe{dominates} $s'_i$\index{dominance} (or
equivalently, that $s'_i$ is
 \bfe{dominated by} $s_i$) if
\[
\fa s_{-i} \in S_{-i} \ p_i(s_i, s_{-i}) \geq p_i(s'_i, s_{-i}).
\]
Further, we say that $s_i$ is \bfe{strictly
  dominant}\index{strategy!strictly dominant} if it strictly dominates
all other strategies of player $i$ and define analogously a
\bfe{weakly dominant}\index{strategy!weakly dominant} and a
\bfe{dominant}\index{strategy!dominant} strategy.

Clearly, a rational player will not choose a strictly dominated
strategy.  As an illustration let us return to the Prisoner's Dilemma.
In this game for each player, $C$ (cooperate) is a strictly dominated
strategy. So the assumption of players' rationality implies that each
player will choose strategy $D$ (defect).  That is, we can predict
that rational players will end up choosing the joint strategy $(D,D)$
in spite of the fact that the Pareto efficient outcome $(C,C)$ yields
for each of them a strictly higher payoff.

The Prisoner's Dilemma game can be easily generalised to $n$ players
as follows.  Assume that each player has two strategies, $C$
and $D$.  Denote by $C^n$ the joint strategy in
which each strategy equals $C$ and similarly with $D^n$.  Further,
given a joint strategy $s_{-i}$ of the opponents of player $i$ denote
by $|s_{-i}(C)|$ the number of $C$ strategies in $s_{-i}$.

Assume now that $k_i$ and $l_i$, where $i \in \{1, \ldots, n\}$, are real numbers such that
for all $i \in \{1, \ldots, n\}$ we have $k_i (n-1) > l_i > 0$.
We put
\[
p_i(s) := \begin{cases}
        k_i |s_{-i}(C)| + l_i     & \mathrm{if}\  s_{i} = D \\
        k_i |s_{-i}(C)|       & \mathrm{if}\  s_{i} = C.
        \end{cases}
\]
Note that for  $n =2, k_i = 2$ and $l_i = 1$ we get the original Prisoner's Dilemma game.

Then for all players $i$ we have $p_i(C^n) = k_i (n-1) > l_i =
p_i(D^n)$, so for all players the strategy profile $C^n$ yields a
strictly higher payoff than $D^n$.  Yet for all players $i$ 
strategy $C$ is strictly dominated by strategy $D$, since
for all $s_{-i} \in S_{-i}$ we have $p_i(D,s_{-i}) - p_i(C,s_{-i}) = l_i > 0$.

Whether a rational player will never choose a weakly dominated strategy is a more subtle
issue that we shall not pursue here.  



By definition, no player achieves a higher payoff by switching from a
dominant strategy to another strategy.  This explains the following
obvious observation.

\begin{note}[Dominant Strategy] \label{not:dominant}
Consider a strategic game $G$.
Suppose that $s$ is a joint strategy such that each
$s_i$ is a dominant strategy.
Then it is a Nash equilibrium of $G$.
\end{note}
%



In particular, the conclusion of the lemma holds if each $s_i$ is a
strictly or a weakly dominant strategy.  In the former case, when the game is finite,
we can additionally assert (see the IESDS Theorem \ref{thm:ies} below)
that $s$ is a unique Nash equilibrium of $G$.  This stronger
claim does not hold if each $s_i$ is a weakly dominant strategy.
Indeed, consider the game
\begin{center}
\begin{game}{2}{2}
      & $L$    & $R$\\
$T$   &$1,1$   &$1,1$\\
$B$   &$1,1$   &$0,0$
\end{game}
\end{center}
Here $T$ is a weakly dominant strategy for the player 1,
$L$ is a weakly dominant strategy for player 2 and,
as prescribed by the above Note,
$(T,L)$, is a Nash equilibrium. However, 
this game has two other Nash equilibria,
$(T,R)$ and~$(B,L)$.

The converse of the above Note of course is not true. Indeed, there are games in
which no strategy is dominant, and yet they have a Nash equilibrium.
An example is the Battle of the Sexes game that has two Nash
equilibria, but no dominant strategy.  


So to find a Nash equilibrium (or Nash equilibria) of a game it does
not suffice to check whether a dominant strategy exists. In what
follows we investigate whether iterated elimination of strategies can
be of help.

\section{Iterated elimination of strategies I}
\label{sec:iter1}

\subsection{Elimination of strictly dominated strategies}
\label{subsec:iterated-strictly}

We assumed that each player is rational. 
So when searching for an outcome
that is optimal for all players we can safely remove strategies that are
strictly dominated by some other strategy.  This can be done in a
number of ways. For example, we could remove all or some strictly
dominated strategies simultaneously, or start removing them in a round
robin fashion starting with, say, player 1.  To discuss this matter
more rigorously we introduce the notion of a restriction of a game.




Given a game $G := (S_1, \ldots, S_n, p_1, \ldots, p_n)$ and (possibly empty) sets
of strategies $R_1, \ldots, R_n$ such that $R_i \sse S_i$ for $i \in
\C{1, \ldots, n}$ we say that $R := (R_1, \ldots, R_n, p_1, \ldots, p_n)$ is a
\bfe{restriction}\index{game!restriction of}
of $G$.  Here of course we view each $p_i$ as a function
on the subset $R_1 \times \ldots \times R_n$ of $S_1 \times \ldots \times S_n$.
In what follows, given a restriction $R$ we denote by $R_i$ the set of strategies
of player $i$ in $R$.

We now introduce the following
notion of reduction between the restrictions $R$ and $R'$ of $G$:
\[
R \myra_{\hspace{-1mm} S \: } R'
\]
when $R \neq R'$, $\fa i \in \C{1, \ldots, n} \ R'_i \sse R_i$ and
\[
\mbox{$\fa i \in \C{1, \ldots, n} \ \fa s_i \in R_i \setminus R'_i \ \te s'_i \in R_i$ $s_i$ is strictly dominated in $R$ by $s'_i$}.
\]
That is, $R \myra_{\hspace{-1mm} S \: } R'$ when $R'$ results from $R$ by removing from it some strictly dominated strategies.

In general an elimination of strictly dominated strategies is not a
one step process; it is an iterative procedure.  Its use is justified
by the assumption of common knowledge of rationality.

\begin{example} \label{exa:2by3}

Consider the following game:

\begin{center}
\begin{game}{3}{3}
      & $L$    & $M$  & $R$\\
$T$   &$3,0$   &$2,1$ &$1,0$\\
$C$   &$2,1$   &$1,1$ &$1,0$\\
$B$   &$0,1$   &$0,1$ &$0,0$
\end{game}
\end{center}

Note that $B$ is strictly dominated by $T$ and
$R$ is strictly dominated by $M$.
By eliminating these two strategies we get:

\begin{center}
\begin{game}{2}{2}
      & $L$    & $M$  \\
$T$   &$3,0$   &$2,1$ \\
$C$   &$2,1$   &$1,1$ \\
\end{game}
\end{center}

Now $C$ is strictly dominated by $T$, so we get:

\begin{center}
\begin{game}{1}{2}
      & $L$    & $M$  \\
$T$   &$3,0$   &$2,1$ \\
\end{game}
\end{center}

In this game $L$ is strictly dominated by $M$, so we finally get:

\begin{center}
\begin{game}{1}{1}
      & $M$  \\
$T$   &$2,1$ \\
\end{game}
\end{center}
\HB
\end{example}

This brings us to the following notion, where given a binary relation $\myra$
we denote by $\myra^{*}$ its transitive reflexive closure.
Consider a strategic game $G$.
Suppose that $G \myra^{*}_{\hspace{-1mm} S} R$, i.e.,  
$R$ is obtained by an iterated elimination of strictly dominated strategies,
in short \bfe{IESDS}, starting with $G$.
\index{IESDS (iterated elimination of strictly dominated strategies)}

\begin{itemize}
  
\item If for no restriction $R'$ of $G$, $R \myra_{\hspace{-1mm} S} R'$ holds,
  we say that $R$ is \bfe{an outcome of IESDS from $G$}.\index{outcome of IESDS}
  
\item If each player is left in $R$ with exactly one strategy, we say
  that $G$ \bfe{is solved by IESDS}.
\end{itemize}

The following simple result clarifies the relation between the IESDS and Nash
equilibrium.

\begin{theorem}[IESDS]
\label{thm:ies}

Suppose that $G'$ is an outcome of IESDS 
from a strategic game $G$.

  \begin{enumerate}[(i)]
  \item If $s$ is a Nash equilibrium of $G$, then it is a Nash equilibrium of $G'$.

  \item If $G$ is finite and $s$ is a Nash equilibrium of $G'$, then it is a Nash equilibrium of $G$.

  \item If $G$ is finite and solved by IESDS, then the resulting joint strategy
    is a unique Nash equilibrium.
  \end{enumerate}
\end{theorem}

\begin{exercise}
  Provide the proof.
\HB
\end{exercise}

\begin{example} \label{exa:loc}
A nice example of a game that is solved by IESDS is the \bfe{location game} due to 
\cite{Hot29}.
Assume that that the players are two vendors who simultaneously choose a location.
Then the customers choose the closest vendor. The profit for each vendor equals the number of
customers it attracted. 

  To be more specific we assume that the vendors choose a location
  from the set $\{1, \ldots, n\}$ of natural numbers, viewed as points
  on a real line, and that at each location there is exactly one
  customer.  For example, for $n = 11$ we have 11 locations: 

\begin{figure}[htbp]
\centering
\input{line.pstex_t}
\end{figure}

\NI
and when the players choose respectively the locations 3 and 8:

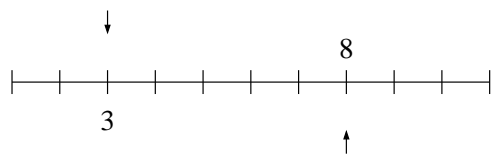
\begin{figure}[htbp]
\centering
\input{line2.pstex_t}
\end{figure}

\NI
we have $p_1(3,8) = 5$ and $p_2(3,8) = 6$.
When the vendors `share' a customer, 
they end up with a fractional payoff.

In general, we  have the following game:

\begin{itemize}
\item each set of strategies consists of the set $\{1, \ldots, n\}$,

\item each payoff function $p_i$ is defined by:
\[
p_i(s_i, s_{3-i}) := \begin{cases} 
\dfrac{s_i + s_{3-i} - 1}{2} &  \mbox{if $s_i < s_{3-i}$} \\[2mm]
n - \dfrac{s_i + s_{3-i} - 1}{2} &  \mbox{if $s_i > s_{3-i}$} \\[2mm]
\dfrac{n}{2} &  \mbox{if $s_i = s_{3-i}$.} 
\end{cases}
\]
\end{itemize}

It is easy to see that for $n = 2 k + 1$ this game is solved by $k$ rounds of IESDS, and that each player
is left with the `middle' strategy $k$. In each round both
`outer' strategies are eliminated, so first 1 and $n$, and so on.
\HB  
\end{example}

There is one more natural question that we left so far unanswered.  Is
the outcome of an iterated elimination of strictly dominated
strategies unique, or in game theory parlance: is strict dominance
\bfe{order independent}?\index{order independence} The answer is positive.  The following
result was established independently by \cite{GKZ90} and \cite{Ste90}.

\begin{theorem}[Order Independence I] \label{thm:strict}
Given a finite strategic game all iterated eliminations of strictly dominated strategies
yield the same outcome.
\end{theorem}

As noted by \cite{DS02} the above result does not hold for infinite
strategic games.  

\begin{example} \label{exa:inf}
Consider a game in which the set of
strategies for each player is the set of natural numbers. The payoff
to each player is the number (strategy) he selected.  

Note that in this game every strategy is strictly dominated.
Consider now three ways of using IESDS:

\begin{itemize}
\item by removing in one step all strategies that are strictly dominated,

\item by removing in one step all strategies different from 0 that are strictly dominated,

\item by removing in each step exactly one strategy.
\end{itemize}

In the first case we obtain the restriction with the empty strategy sets,
in the second one we end up with the restriction in which each player has
just one strategy, 0, and in the third case we obtain
an infinite sequence of reductions. 
\HB
\end{example}

The above example shows that in the limit of an infinite sequence of
reductions different outcomes can be reached.  So for infinite games
the definition of the order independence has to be modified.  An
interested reader is referred to \cite{DS02} and \cite{Apt07} where
two different options are proposed and some limited order independence
results are established.

The above example also shows that in the IESDS Theorem
\ref{thm:ies}$(ii)$ and $(iii)$ we cannot drop the assumption that the
game is finite.  Indeed, the above infinite game has no Nash
equilibria, while the game in which each player has exactly one
strategy has a Nash equilibrium.

\subsection{Elimination of weakly dominated strategies}
\label{subsec:iterated-weakly}

Analogous considerations can be carried out for the elimination of
weakly dominated strategies, by considering the appropriate reduction relation
$\myra_{\hspace{-1mm} W}$ 
defined in the expected way.  
Below we abbreviate iterated elimination of weakly dominated strategies
to \bfe{IEWDS}.
\index{IEWDS (iterated elimination of weakly dominated strategies)}

However, in the case of IEWDS some complications arise.  To illustrate
them consider the following game that results from equipping each
player in the Matching Pennies game with a third strategy $E$ (for
Edge):

\begin{center}
\begin{game}{3}{3}
      & $H$     & $T$       & $E$ \\
$H$   & $\phantom{-}1,-1$  & $-1,\phantom{-}1$    & $-1,-1$ \\
$T$   & $-1,\phantom{-}1$  & $\phantom{-}1,-1$    & $-1,-1$ \\
$E$   & $-1,-1$            & $-1,-1$              & $-1,-1$ \\
\end{game}
\end{center}

Note that 

\begin{itemize}
\item $(E,E)$ is its only Nash equilibrium,

\item for each player, $E$ is the only strategy that is
weakly dominated.
\end{itemize}

Any form of elimination of these two $E$ strategies, simultaneous or
iterated, yields the same outcome, namely the Matching Pennies game,
that, as we have already noticed, has no Nash equilibrium.  So during
this eliminating process we `lost' the only Nash equilibrium.  In
other words, part $(i)$ of the IESDS Theorem
\ref{thm:ies} does not hold when reformulated for weak
dominance. 

On the other hand, some partial results are still valid here.





\begin{theorem}[IEWDS]
\label{thm:iter2}

Suppose that $G$ is a finite strategic game.

  \begin{enumerate}[(i)]
  \item If $G'$ is an outcome of IEWDS from $G$ and $s$ is a Nash equilibrium of $G'$, then
$s$ is a Nash equilibrium of $G$.

  \item If $G$ is solved by IEWDS, then the resulting joint strategy is 
    a Nash equilibrium of $G$. 
  \end{enumerate}
\end{theorem}
\begin{exercise}
  Provide the proof.
\HB
\end{exercise}

\begin{example}\label{exa:beauty}
A nice example of a game that is solved by IEWDS is the \bfe{Beauty Contest game} due to 
\cite{Mou86}. In this game there are $n > 2$ players, each with the set of strategies equal $\{1, \ldots, 100\}$.
Each player submits a number and the payoff to each player is obtained by splitting 1
equally between the players whose submitted number is closest to $\frac{2}{3}$ of the average.
For example, if the submissions are $29,32,29$, then the payoffs are respectively $\frac{1}{2}, 0, \frac{1}{2}$.

One can check that this game is solved by IEWDS and results in the
joint strategy $(1, \ldots, 1)$.  Hence, by the IEWDS Theorem
\ref{thm:iter2} this joint strategy is a (not necessarily unique; we
shall return to this question in Section \ref{sec:iter2}) Nash
equilibrium.
\HB
\end{example}
\begin{exercise}
  Show that the Beauty Contest game is indeed solved by IEWDS.

\HB
\end{exercise}

Note that in contrast to the IESDS Theorem \ref{thm:ies} we do not
claim in part $(ii)$ of the  IEWDS Theorem \ref{thm:iter2}
that the resulting joint strategy is a \emph{unique} Nash equilibrium.
In fact, such a stronger claim does not hold. Further, in contrast to
strict dominance, an iterated elimination of weakly dominated
strategies can yield several outcomes.

The following example reveals even more peculiarities of this procedure.

\begin{example} \label{exa:wd}
Consider the following game:

 \begin{center}
 \begin{game}{2}{3}
       & $L$   & $M$     & $R$  \\
 $T$   &$0,1$  & $1,0$   & $0,0$  \\
 $B$   &$0,0$  & $0,0$   & $1,0$  
 \end{game}
 \end{center}
It has three Nash equilibria, $(T,L)$, $(B,L)$ and $(B,R)$.
This game can be solved by IEWDS but only if in the first round we do not
eliminate all weakly dominated strategies, which are $M$ and $R$. 
If we eliminate only $R$, then we reach the game

 \begin{center}
 \begin{game}{2}{2}
       & $L$   & $M$  \\
 $T$   &$0,1$  & $1,0$  \\
 $B$   &$0,0$  & $0,0$ 
 \end{game}
 \end{center}
that is solved by IEWDS by eliminating $B$ and $M$. This yields
 \begin{center}
 \begin{game}{1}{1}
       & $L$ \\
 $T$   &$0,1$ 
 \end{game}
 \end{center}
So not only IEWDS is not order independent; in some games it is advantageous \emph{not} to proceed
with the deletion of the weakly dominated strategies `at full speed'.
The reader may also check that the second Nash equilibrium, $(B,L)$, can be found using IEWDS, as well,
but not the third one, $(B,R)$.

\HB  
\end{example}



To summarise, the iterated elimination of weakly dominated strategies

\begin{itemize}
\item can lead to a deletion of Nash equilibria, 

\item does not need to yield a unique outcome,

\item can be too restrictive if we stipulate that in each round all weakly dominated strategies are eliminated.

\end{itemize}

Finally, note that the
above IEWDS Theorem \ref{thm:iter2} does not hold for infinite games.
Indeed, Example \ref{exa:inf} applies here, as well.






\subsection{Elimination of never best responses}
\label{subsec:iterated-nbr}

Finally, we consider the process of eliminating strategies that are never best responses to
a joint strategy of the opponents. 
To motivate this procedure consider the following game:
 \begin{center}
 \begin{game}{3}{2}
       & $X$   & $Y$  \\
 $A$   &$2,1$  & $0,0$  \\
 $B$   &$0,1$  & $2,0$  \\
 $C$   &$1,1$  & $1,2$  
 \end{game}
 \end{center}

Here no strategy is strictly or weakly dominated. However, 
$C$ is a \bfei{never best response}, that is, it is not a best response to any 
strategy of the opponent. Indeed, $A$ is a unique best response to $X$ and $B$ is a
unique best response to $Y$.
Clearly, the above game is
solved by an iterated elimination of never best responses. So this procedure
can be stronger than IESDS and IEWDS.

Formally, we introduce the following reduction notion between the restrictions $R$ and $R'$ of a given strategic game $G$:
\[
R \myra_{\hspace{-1mm} N \: } R'
\]
when $R \neq R'$, $\fa i \in \C{1, \ldots, n} \ R'_i \sse R_i$ and
\[
\mbox{$\fa i \in \C{1, \ldots, n} \ \fa s_i \in R_i \setminus R'_i \ \neg \te s_{-i} \in R_{-i}$ $s_i$ is a best response to $s_{-i}$ in $R$}.
\]
That is, $R \myra_{\hspace{-1mm} N \: } R'$ when $R'$ results from $R$ by removing from it some strategies 
that are never best responses.

We then focus on the iterated elimination of never best responses, in
short \bfe{IENBR}, \index{IENBR (iterated elimination of never best
  responses)} obtained by using the $\myra^{*}_{\hspace{-1mm} N}$
relation.  The following counterpart of the IESDS Theorem
\ref{thm:ies} then holds.

\begin{theorem}[IENBR]
\label{thm:nbr}

Suppose that $G'$ is an outcome of IENBR
from a strategic game $G$.

  \begin{enumerate}[(i)]
  \item If $s$ is a Nash equilibrium of $G$, then it is a Nash equilibrium of $G'$.

  \item If $G$ is finite and $s$ is a Nash equilibrium of $G'$, then it is a Nash equilibrium of $G$.
    
  \item If $G$ is finite and solved by IENBR, then the resulting joint strategy
    is a unique Nash equilibrium.
  \end{enumerate}
\end{theorem}
\begin{exercise}
  Provide the proof.
\HB
\end{exercise}

Further, as shown by \cite{Apt05}, we have the following analogue of 
the Order Independence I Theorem \ref{thm:strict}.

\begin{theorem}[Order Independence II] \label{thm:strict2}
Given a finite strategic game all iterated eliminations of never best responses
yield the same outcome.
\end{theorem}

In the case of infinite games we encounter the same problems as in the case of IESDS
as Example \ref{exa:inf} readily applies to IENBR, as well.
In particular, if we solve an infinite game by IENBR we cannot claim that 
we obtained a Nash equilibrium. Still, IENBR can be useful in such cases.

\begin{example} \label{exa:loc1}
  Consider the following infinite variant of the location game considered in Example \ref{exa:loc}.
We assume that the players choose their strategies from the open interval $(0,100)$ and that
at each real in $(0,100)$ there resides one customer. We have then
the following payoffs that correspond to the intuition that the customers choose the closest vendor:
\[
p_i(s_i, s_{3-i}) := \begin{cases} 
\dfrac{s_i + s_{3-i}}{2} &  \mbox{if $s_i < s_{3-i}$} \\[2mm]
100 - \dfrac{s_i + s_{3-i}}{2} &  \mbox{if $s_i > s_{3-i}$} \\[2mm]
50 &  \mbox{if $s_i = s_{3-i}$.} 
\end{cases}
\]

It is easy to check that in this game no strategy strictly or weakly dominates another one.
On the other hand each strategy 50 is a best response to some strategy, namely to 50,
and no other strategies are best responses. So this game is solved by IENBR, in one step.
We cannot claim automatically that the resulting joint strategy $(50,50)$ is a Nash equilibrium, but
it is straightforward to check that this is the case. Moreover, by 
the IENBR Theorem \ref{thm:nbr}$(i)$ we know that this is a unique Nash equilibrium.
\HB
\end{example}

\section{Mixed extension}

We now study a special case of infinite strategic games that are obtained in a canonical
way from the finite games, by allowing mixed strategies.
Below $[0,1]$ stands for the real interval $\{r \in \mathbb{R} \mid 0 \leq r \leq 1\}$.
By a \bfei{probability distribution} over a finite non-empty set $A$ we mean a function
\[
\pi: A \myra [0,1]
\]
such that $\sum_{a \in A} \pi(a) = 1$. We denote the set of probability distributions over
$A$ by $\Delta A$.

Consider now a finite strategic game $G := (S_1, \ldots, S_n, p_1, \ldots, p_n)$.
By a \bfe{mixed strategy}\index{strategy!mixed}
 of player $i$ in $G$ we mean a probability distribution over $S_i$. So
$\Delta S_i$ is the set of mixed strategies available to player $i$.
In what follows, we denote a mixed strategy of player $i$ by $m_i$ and 
a joint mixed strategy of the players by $m$.

Given a mixed strategy $m_i$ of player $i$ we define
\[
support(m_i) := \{a \in S_i \mid m_i(a) > 0\}
\]
and call this set the \bfe{support}\index{strategy!support of}
 of $m_i$.
In specific examples we write a mixed strategy $m_i$
as the sum $\sum_{a \in A} m_i(a) \cdot a$, where $A$ is the support of $m_i$.

Note that in contrast to $S_i$ the set $\Delta S_i$ is infinite.  When
referring to the mixed strategies, as in the previous sections, we use
the `$_{-i}$' notation. So for $m \in \Delta S_1 \times \ldots \times
\Delta S_n$ we have
$m_{-i} = (m_j)_{j \neq i}$, etc.

We can identify each strategy $s_i \in S_i$ with the mixed strategy 
that puts `all the weight' on the strategy $s_i$.
In this context $s_i$ will be called a \bfe{pure strategy}.\index{strategy!pure}
Consequently we can view $S_i$ as a subset of $\Delta S_i$ and
$S_{-i}$ as a subset of $\times_{j \neq i} \Delta  S_j$.

By a \bfe{mixed extension}\index{game!mixed extension of} of 
$(S_1, \ldots, S_n, p_1, \ldots, p_n)$ we mean the strategic game
\[
(\Delta S_1, \ldots, \Delta S_n, p_1, \ldots, p_n),
\]
where each function $p_i$ is extended in a canonical way
from 
$S := S_1 \times \ldots \times S_n$
to $M := \Delta S_1 \times \ldots \times \Delta S_n$
by first viewing each joint mixed strategy
$m = (m_1, \ldots, m_n) \in M$ as a probability distribution over $S$, 
by putting for $s \in S$
\[
m(s) := m_1(s_1) \cdot \ldots \cdot m_n(s_n),
\]
and then by putting
\[
p_i(m) := \sum_{s \in S} m(s) \cdot p_i(s).
\]

The notion of a Nash equilibrium readily applies to mixed extensions.
In this context we talk about a \bfe{pure Nash equilibrium},
\index{Nash equilibrium!pure} when each of the constituent strategies is pure,
and refer to an arbitrary Nash equilibrium of the mixed extension as a
\bfe{Nash equilibrium in mixed strategies}\index{Nash equilibrium!in
  mixed strategies} of the initial finite game.  In what follows, when
we use the letter $m$ we implicitly refer to the latter Nash
equilibrium.












  

\begin{lemma}[Characterisation] \label{lem:characterization}

Consider a finite strategic game 

\NI
$(S_1, \ldots, S_n, p_1, \ldots, p_n)$.
The following statements are equivalent:

\begin{enumerate}[(i)]
\item $m$ is a Nash equilibrium in mixed strategies, i.e.,
\[
p_i(m) \geq p_i(m'_i, m_{-i})
\]
for all $i \in \C{1, \ldots, n}$ and all $m'_i \in \Delta S_i$,

\item for all $i \in \C{1, \ldots, n}$ and all $s_i \in S_i$
\[
p_{i}(m) \geq p_i(s_i, m_{-i}),
\]

\item for all $i \in \C{1, \ldots, n}$ and all $s_i \in support(m_i)$
\[
p_i(m) = p_i(s_i, m_{-i})
\]
and for all $i \in \C{1, \ldots, n}$ and all $s_i \not\in support(m_i)$
\[
p_i(m) \geq p_i(s_i, m_{-i}).
\]
\end{enumerate}
\end{lemma}
\begin{exercise}
  Provide the proof.
\HB
\end{exercise}

Note that the equivalence between $(i)$ and $(ii)$ implies that each
Nash equilibrium of the initial game is a pure Nash equilibrium of
the mixed extension.  In turn, the equivalence between $(i)$ and
$(iii)$ provides us with a straightforward way of testing whether a
joint mixed strategy is a Nash equilibrium.






We now illustrate the use of the above theorem by finding in the
Battle of the Sexes game a Nash equilibrium in mixed strategies, in
addition to the two pure ones exhibited in Section \ref{sec:iter1}.
Take
\[
  \begin{array}{l}
m_1 := r_1 \cdot F + (1 - r_1) \cdot B, \\
m_2 := r_2 \cdot F + (1 - r_2) \cdot B,
  \end{array}
\]
where $0 < r_1, r_2 < 1$. By definition
\[
  \begin{array}{l}
p_1(m_1, m_2) = 2 \cdot r_1 \cdot r_2 + (1 - r_1) \cdot (1 - r_2), \\
p_2(m_1, m_2) = r_1 \cdot r_2 + 2 \cdot (1 - r_1) \cdot (1 - r_2). \\
  \end{array}
\]

Suppose now that $(m_1, m_2)$ is a Nash equilibrium in mixed
strategies.  By the equivalence between $(i)$ and $(iii)$ of the
Characterisation Lemma \ref{lem:characterization} $p_1(F, m_2) =
p_1(B, m_2)$, i.e., (using $r_1 = 1$ and $r_1 = 0$ in the above
formula for $p_1(\cdot)$) $2 \cdot r_2 = 1 - r_2$, and $p_2(m_1, F) =
p_2(m_1, B)$, i.e., (using $r_2 = 1$ and $r_2 = 0$ in the above
formula for $p_2(\cdot)$) $r_1 = 2 \cdot (1 - r_1)$.  So $r_2 =
\frac{1}{3}$ and $r_1 = \frac{2}{3}$.

This implies that for these values of $r_1$ and $r_2$,
$(m_1, m_2)$ is a Nash equilibrium in mixed strategies and we have
\[
  \begin{array}{l}
p_1(m_1, m_2) = p_2(m_1, m_2) = \frac{2}{3}.
  \end{array}
\]

The example of the Matching Pennies game illustrated that some
strategic games do not have a Nash equilibrium.  In the case of
mixed extensions the situation changes and we have the following
fundamental result due to \cite{Nas50}.

\begin{theorem}[Nash] \label{thm:nash} \index{theorem!Nash}%
  Every mixed extension of a finite strategic game has a Nash equilibrium.
\end{theorem}

In other words, every finite strategic game has a Nash equilibrium in
mixed strategies.  In the case of the Matching Pennies game it is
straightforward to check that $(\frac{1}{2} \cdot H + \frac{1}{2}
\cdot T, \frac{1}{2} \cdot H + \frac{1}{2} \cdot T)$ is such a Nash
equilibrium.  In this equilibrium the payoffs to each player are 0.

Nash's Theorem follows directly from the following result due to
\cite{Kak41}.\footnote{Recall that a subset $A$ of $\mathbb{R}^{n}$ is
  called \bfe{compact}\index{set!compact}
 if it is closed and bounded, and is called
  \bfe{convex}\index{set!convex} if for any $\textbf{x}, \textbf{y} \in A$ and
  $\alpha \in [0,1]$ we have $\alpha \textbf{x} + (1 - \alpha)
  \textbf{y} \in A$.}

\begin{theorem}[Kakutani]
Suppose that $A$ is a non-empty compact and convex subset of
${\cal R}^n$ and
\[
\Phi: A \myra {\cal P}(A)
\]
such that

\begin{itemize}
\item $\Phi(x)$ is non-empty and convex for all $x \in A$,

\item the {\begin{bfseries}\textit{graph}\end{bfseries}}
of $\Phi$, so the set $\{(x,y) \mid y \in \Phi(x)\}$,
is closed.
\end{itemize}
Then $x^{*} \in A$ exists such that $x^{*} \in \Phi(x^{*})$.
\HB
\end{theorem}

\II

\NI
\emph{Proof of Nash's Theorem}.
Fix a finite strategic game $(S_1, \ldots, S_n, p_1, \ldots, p_n)$.
Define the function
$
best_i : \times_{j \neq i} \Delta  S_j \myra {\cal P}(\Delta S_i)
$
by
\[
best_i(m_{-i}) := \{m_i \in \Delta S_i \mid m_i \mbox{ is a best response to } m_{-i}\}.
\]
Then define the function
$
best : \Delta S_1 \times \ldots \times \Delta S_n \myra {\cal P}(\Delta S_1 \times \ldots \times \Delta S_n)
$
by
\[
best(m) := best_1(m_{-1}) \times \ldots \times best_1(m_{-n}).
\]

It is now straightforward to check that $m$ is a Nash equilibrium iff
$m \in best(m)$.  Moreover, one can easily check that the function
$best(\cdot)$ satisfies the conditions of Kakutani's
Theorem. The fact that for every joint mixed strategy $m$,
  $best(m)$ is non-empty is a direct consequence of the Extreme Value
  Theorem stating that every real-valued continuous function on a
  compact subset of ${\cal R}^{\ell}$ attains a maximum.  \HB \III

\section{Iterated elimination of strategies II}
\label{sec:iter2}

The notions of dominance apply in particular to mixed extensions of
finite strategic games. But we can also consider dominance of a
\emph{pure} strategy by a \emph{mixed} strategy.  Given a finite
strategic game $G := (S_1, \ldots, S_n, p_1, \ldots, p_n)$, we say that a
(pure) strategy $s_i$ of player $i$ is \bfe{strictly dominated by}
a mixed strategy $m_i$ if
\[
\fa s_{-i} \in S_{-i} \ p_i(m_i, s_{-i}) > p_i(s_i, s_{-i}),
\]
and that $s_i$ is \bfe{weakly dominated by} a mixed strategy $m_i$ if
\[
\mbox{$\fa s_{-i} \in S_{-i} \ p_i(m_i, s_{-i}) \geq p_i(s_i, s_{-i})$ and
$\te s_{-i} \in S_{-i} \ p_i(m_i, s_{-i}) > p_i(s_i, s_{-i})$}.
\]

In what follows we discuss for these two forms of
dominance the counterparts of the results presented
in Section \ref{sec:iter1}.

\subsection{Elimination of strictly dominated strategies}
\label{subsec:iterated-strictlyII}

Strict dominance by a mixed strategy leads to a stronger notion of
strategy elimination.  For example, in the game
\begin{center}
\begin{game}{3}{2}
      & $L$    & $R$\\
$T$   &$2,1$   &$0,1$\\
$M$   &$0,1$   &$2,1$\\
$B$   &$0,1$   &$0,1$
\end{game}
\end{center}
the strategy $B$ is strictly dominated neither by $T$ nor $M$ but is
strictly dominated by $\frac{1}{2} \cdot T + \frac{1}{2} \cdot M$.

We now focus on iterated elimination of pure strategies that are
strictly dominated by a mixed strategy. As in Section \ref{sec:iter1}
we would like to clarify whether it affects the Nash equilibria,
in this case equilibria in mixed strategies.

Instead of the lengthy wording `the iterated elimination of strategies
strict\-ly dominated by a mixed strategy' we write \bfei{IESDMS}.  We
have then the following counterpart of the IESDS Theorem
\ref{thm:ies}, where we refer to Nash equilibria in mixed strategies.
Given a restriction $G'$ of $G$ and a joint mixed
strategy $m$ of $G$, when we say that $m$ is a Nash equilibrium of
$G'$ we implicitly stipulate that each strategy used (with positive
probability) in $m$ is a strategy in $G'$.

\begin{theorem}[IESDMS]
\label{thm:imes}

Suppose that $G$ is a finite strategic game.

  \begin{enumerate}[(i)]
  \item If $G'$ is an outcome of IESDMS from $G$, then $m$ is a Nash
    equilibrium of $G$ iff it is a Nash equilibrium of $G'$.
    
  \item If $G$ is solved by IESDMS, then the resulting joint strategy is a
    unique Nash equilibrium of $G$  (in, possibly, mixed strategies).  
  \end{enumerate}
\end{theorem}
\begin{exercise}
  Provide the proof.
\HB
\end{exercise}

To illustrate the use of this result let us return to the Beauty Contest game
discussed in Example \ref{exa:beauty}. We explained there why $(1,
\ldots, 1)$ is \emph{a} Nash equilibrium. Now we can draw a stronger conclusion.

\begin{example} 
  One can show that the Beauty Contest game is solved by IESDMS in 99 rounds. In
  each round the highest strategy of each player is removed and
  eventually each player is left with the strategy 1. On account
  of the above theorem we now conclude that $(1, \ldots, 1)$ is a
  \emph{unique} Nash equilibrium.
\HB
\end{example}

\begin{exercise}
   Show that the Beauty Contest game is indeed solved by IESDMS in 99 rounds.
\HB
\end{exercise}

As in the case of strict dominance by a pure strategy we now address
the question of whether the outcome of IESDMS is unique.  The answer, as
before, is positive.  The following result was established by
\cite{OR94}.

\begin{theorem}[Order independence III] \label{thm:strictmixed}
All iterated eliminations of strat\-egies strictly dominated by a mixed strategy
yield the same outcome.
\end{theorem}

\subsection{Elimination of weakly dominated strategies}
\label{subsec:iter-weak-mixed}

Next, we consider iterated elimination of pure strategies that are
weakly dominated by a mixed strategy.

As already noticed in Subsection \ref{subsec:iterated-weakly} an
elimination by means of weakly dominated strategies can result in a
loss of Nash equilibria.  Clearly, the same observation applies here.
%
%
%
%
%
We also have the following counterpart of the IEWDS Theorem
\ref{thm:iter2}, where we refer to Nash equilibria in mixed strategies.
Instead of `the iterated elimination of strategies
weakly dominated by a mixed strategy' we write \bfei{IEWDMS}.

\begin{theorem}[IEWDMS]
\label{thm:imewds}

Suppose that $G$ is a finite strategic game.

  \begin{enumerate}[(i)]
  \item If $G'$ is an outcome of IEWDMS from $G$ and 
$m$ is a Nash equilibrium of $G'$, then
$m$ is a Nash equilibrium of $G$.

  \item If $G$ is solved by IEWDMS, then the resulting joint strategy is 
    a Nash equilibrium of $G$. 
  \end{enumerate}
\end{theorem}

Here is a simple application of this theorem.

\begin{corollary}
Every mixed extension of a finite strategic game has a Nash equilibrium such that
no strategy used in it is weakly dominated by a mixed strategy.
\end{corollary}

\Proof
It suffices to apply Nash's Theorem \ref{thm:nash} to an
outcome of IEWDMS and use item $(i)$ of the above theorem.
\HB
\III

Finally, observe that the outcome of IEWMDS does not need to be unique. In fact,
Example \ref{exa:wd} applies here, as well.







\subsection{Rationalizability}

Finally, we consider iterated elimination of strategies that are never best responses to 
a joint mixed strategy of the opponents.
Following \cite{Ber84} and \cite{Pea84},
strategies that survive such an elimination  process
are called rationalizable strategies.\footnote{More precisely, in each of these papers
a different definition is used; see \cite{Apt07} for an analysis of the conditions for which
these definitions coincide.}

Formally, we define rationalizable strategies as follows.
Consider a restriction $R$ of a finite strategic game $G$. Let
\[
{\cal RAT}(R) := (S'_1, \ldots, S'_n),
\]
where for all $i \in \{1, \ldots, n\}$
\[
S'_i := \{ s_i \in R_i \mid \te m_{-i} \in \times_{j \neq i} \Delta R_{j} \ \mbox{$s_i$ is a best response to $m_{-i}$ in $G$}
\}.
\]
Note the use of $G$ instead of $R$ in the definition of $S'_i$. We shall comment on it below.

Consider now the outcome $G_{\cal RAT}$ of
iterating ${\cal RAT}$ starting with $G$.  We call then the strategies
present in the restriction $G_{\cal RAT}$ \bfe{rationalizable}.\index{strategy!rationalizable}

We have the following counterpart of the IESDMS Theorem \ref{thm:imes}, 
due to \cite{Ber84}.
\begin{theorem}
\label{thm:rat}

Assume a finite strategic game $G$.
  \begin{enumerate}[(i)]
  \item Then $m$ is a Nash equilibrium of $G$ iff it is a Nash equilibrium of $G_{\cal RAT}$.
    
  \item If each player has in $G_{\cal RAT}$ exactly one strategy, then the
resulting joint strategy is a
    unique Nash equilibrium of $G$.  
  \end{enumerate}
\end{theorem}
\begin{exercise}
  Provide the proof.
\HB
\end{exercise}
In the context of rationalizability a joint mixed strategy of the
opponents is referred to as a \bfei{belief}.  The definition of
rationalizability is generic in the class of beliefs
w.r.t.~which best responses are collected.  For example, 
we could use here joint pure strategies of the opponents, or
probability distributions over the Cartesian product of the opponents'
strategy sets, so the elements of the set $\Delta S_{-i}$ (extending in an
expected way the payoff functions).  In the first case we talk about
\bfe{point beliefs}\index{belief!point} and in the second case about \bfe{correlated
  beliefs}.\index{belief!correlated}

In the case of point beliefs we can apply the elimination procedure entailed
by ${\cal RAT}$ to arbitrary games. 
To avoid discussion of the outcomes reached in the case of infinite iterations we focus
on a result for a limited case. We refer here to Nash equilibria in pure strategies.

\begin{theorem}
\label{thm:rat1}
Assume a strategic game $G$.
Consider the definition of the ${\cal RAT}$ operator for the case of point beliefs
and suppose that the outcome $G_{\cal RAT}$ is reached in finitely many steps.

\begin{enumerate}[(i)]
  \item Then $s$ is a Nash equilibrium of $G$ iff it is a Nash equilibrium of $G_{\cal RAT}$.
    
  \item If each player is left in $G_{\cal RAT}$ with exactly one strategy, then the
resulting joint strategy is a unique Nash equilibrium of $G$.  
  \end{enumerate}
\end{theorem}
\begin{exercise}
  Provide the proof.
\HB
\end{exercise}

A subtle point is that when $G$ is infinite, the restriction $G_{\cal
  RAT}$ may have empty strategy sets (and hence no joint
strategy).

\begin{example} \label{exa:ber}
  
  \bfei{Bertrand competition}, originally proposed by
\cite{Ber83}, is
  a game concerned with a simultaneous selection of prices for the
  same product by two firms. The product is then sold by the firm that
  chose a lower price.  In the case of a tie the product is sold by
  both firms and the profits are split.
  
  Consider a version in which the range of possible prices is the
  left-open real interval $(0, 100]$ and the demand equals $100 - p$,
  where $p$ is the lower price.  So in this game $G$ there are two
  players, each with the set $(0, 100]$ of strategies and the
  payoff functions are defined by:

\[
\begin{array}{l}
p_1(s_1, s_2) := \begin{cases}
s_1 (100 - s_1) &  \mbox{if $s_1 < s_2$} \\[2mm]
\dfrac{s_1 (100 - s_1)}{2} &  \mbox{if $s_1 = s_2$} \\[2mm]
0 &  \mbox{if $s_1 > s_2$} 
\end{cases}
\\
\\
p_2(s_1, s_2) := \begin{cases} 
s_2 (100 - s_2) &  \mbox{if $s_2 < s_1$} \\[2mm]
\dfrac{s_2 (100 - s_2)}{2} &  \mbox{if $s_1 = s_2$} \\[2mm]
0 &  \mbox{if $s_2 > s_1$.} 
\end{cases}
\end{array}
\]


Consider now each player's best responses to the strategies of the
opponent.  Since $s_1 = 50$ maximises the value of $s_1 (100 - s_1)$
in the interval $(0, 100]$, the strategy 50 is the unique best
response of the first player to any strategy $s_2 > 50$ of the second
player.  Further, no strategy is a best response to a strategy $s_2
\leq 50$.  By symmetry the same holds for the strategies of the second
player.

So the elimination of never best responses leaves each player with a single
strategy, 50. 
In the second round we need to consider the best
responses to these two strategies in the \emph{original} game $G$. In
$G$ the strategy $s_1 = 49$ is a better response to $s_2 = 50$ than
$s_1 = 50$ and symmetrically for the second player.  So in the second
round of elimination both strategies 50 are eliminated and we reach
the restriction with the empty strategy sets.  By Theorem
\ref{thm:rat1} we conclude that the original game $G$ has no Nash
equilibrium.  

\HB
\end{example}

Note that if we defined $S'_i$ in the definition of the operator $\cal
RAT$ using the restriction $R$ instead of the original game $G$, the
iteration would stop in the above example after the first round.  Such
a modified definition of the ${\cal RAT}$ operator is actually an
instance of the IENBR (iterated elimination of never best responses)
in which at each stage all never best responses are eliminated.  So
for the above game $G$ we can then conclude by the IENBR Theorem
\ref{thm:nbr}$(i)$ that it has at most one equilibrium, namely
$(50,50)$, and then check separately that in fact it is not a Nash
equilibrium.

\subsection{A comparison between the introduced notions}

We introduced so far the notions of strict dominance, weak dominance,
and a best response, and related them to the notion of a Nash
equilibrium.  To conclude this section we clarify the connections
between the notions of dominance and of best response.

Clearly, if a strategy is strictly dominated, then it is a never best
response.  However, the converse fails. Further, there is no relation
between the notions of weak dominance and never best response. Indeed,
in the game considered in Subsection \ref{subsec:iterated-nbr}
strategy $C$ is a never best response, yet it is neither strictly nor
weakly dominated.  Further, in the game given in Example \ref{exa:wd}
strategy $M$ is weakly dominated and is also a best response to $B$.

The situation changes in the case of mixed extensions of two-player
finite games.  Below, by a \bfe{totally mixed strategy}\index{strategy!totally mixed} we mean a
mixed strategy with full support, i.e., one in which each strategy is
used with a strictly positive probability.  The following results were
established by \cite{Pea84}. 

\begin{theorem} \label{thm:pea}
  Consider a finite two-player strategic game.
  \begin{enumerate}[(i)]
  \item A pure strategy is strictly dominated by a mixed strategy iff it is not a best response
to a mixed strategy.
    
  \item A pure strategy is weakly dominated by a mixed strategy iff it is not a best response
to a totally mixed strategy.

  \end{enumerate}
\end{theorem}

We only prove here part $(i)$. \cite{Pea84} provides a short, but a
bit tricky proof based on Nash's Theorem \ref{thm:nash}. The proof we
provide, due to \cite{FT91}, is a bit more intuitive.  

We shall use the following result, see, e.g., \cite{Roc70}.

\begin{theorem}[Separating Hyperplane] \label{thm:separating}
  Let $A$ and $B$ be disjoint convex subsets of $\mathbb{R}^k$. 
Then there exists a non-zero $c \in \mathbb{R}^k$ and $d \in \mathbb{R}$ such that
\[
\mbox{$c \cdot x \geq d$ for all $x \in A$,}
\]
\[
\mbox{$c \cdot y \leq d$ for all $y \in B$.}
\]
\end{theorem}
\II

\NI
\emph{Proof of Theorem \ref{thm:pea}$(i)$}.

Clearly,  if a pure strategy is strictly dominated 
by a mixed strategy, then  it is not a best response to a mixed strategy.
To prove the converse, fix a two-player strategic game $(S_1, S_2, p_1, p_2)$.
Also fix $i \in \{1, 2\}$ and abbreviate $3-i$ to $-i$.

Suppose that a strategy $s_i \in S_i$ is not
strictly dominated by a mixed strategy.
Let 
\[
A := \C{x \in \mathbb{R}^{|S_{-i}|} \mid \fa s_{-i} \in S_{-i} \ x_{s_{-i}} > 0}
\]
and
\[
B:= \C{(p_i(m_i, s_{-i})-p_i(s_i, s_{-i}))_{s_{-i} \in S_{-i}} \mid m_i \in \Delta S_i}.
\]
By the choice of $s_i$ the sets $A$ and $B$ are disjoint. Moreover,
both sets are convex subsets of $\mathbb{R}^{|S_{-i}|}$.

By the Separating Hyperplane Theorem \ref{thm:separating}
for some non-zero $c \in \mathbb{R}^{|S_{-i}|}$ and $d \in \mathbb{R}$
\begin{equation}
  \label{equ:geq}
\mbox{$c \cdot x \geq d$ for all $x \in A$,}  
\end{equation}
\begin{equation}
  \label{equ:leq}
\mbox{$c \cdot y \leq d$ for all $y \in B$.}
\end{equation}

But $\textbf{0} \in B$, so by (\ref{equ:leq}) $d \geq 0$.
Hence by (\ref{equ:geq}) and the definition of $A$ for all ${s_{-i}}
\in S_{-i}$ we have $c_{s_{-i}} \geq 0$. Again by (\ref{equ:geq}) 
and the definition of $A$ this
excludes the contingency that $d > 0$, i.e., $d = 0$.  Hence by
(\ref{equ:leq})
\begin{equation}
\mbox{$\sum_{s_{-i} \in S_{-i}} c_{s_{-i}} p_i(m_i, s_{-i}) \leq \sum_{s_{-i} \in S_{-i}} c_{s_{-i}} p_i(s_i, s_{-i})$
for all $m_i \in \Delta S_i$.}
\label{equ:leq2}
\end{equation}

Let $\bar{c} := \sum_{s_{-i} \in S_{-i}} c_{s_{-i}}$. By the assumption $\bar{c} \neq 0$.
Take
\[
m_{-i} := \sum_{s_{-i} \in S_{-i}} \frac{c_{s_{-i}}}{\bar{c}} s_{-i}.
\]
Then (\ref{equ:leq2}) can be rewritten as
\[
\mbox{$p_i(m_i, m_{-i}) \leq p_i(s_i, m_{-i})$ for all $m_i \in \Delta S_i$,}
\]
i.e., $s_i$ is a best response to $m_{-i}$.
\HB

\section{Variations on the definition of strategic games}
\label{sec:variations}

The notion of a strategic game is quantitative in the sense that it refers through
payoffs to real numbers.
A natural question to ask is: do the payoff values matter?  The answer
depends on which concepts we want to study.  We mention here
three qualitative variants of the definition of a strategic game in
which the payoffs are replaced by preferences.  By a
\bfei{preference relation} on a set $A$ we mean here a linear order on $A$.

In \cite{OR94} a strategic game is defined as a sequence 
\[
(S_1, \ldots, S_n, {\succeq}_1, \ldots, {\succeq}_n),
\] 
where each $\succeq_i$ is player's $i$
\bfei{preference relation} defined on the set $S_1
\times \dots \times S_n$ of joint strategies.

In \cite{ARV08} another modification of strategic games is
considered, called a \bfe{strategic game with parametrised
  preferences}.\index{game!strategic with parametrised preferences}
In this approach each player $i$ has a non-empty set of
strategies $S_i$ and a \bfei{preference relation} $\succeq_{s_{-i}}$
on $S_i$ \emph{parametrised} by a joint
strategy $s_{-i}$ of his opponents.
In \cite{ARV08} only strict preferences were considered and so defined
finite games with parametrised preferences were compared with the concept of 
\bfe{CP-nets}\index{CP-net}
 (Conditional Preference nets), a
formalism used for representing conditional and qualitative preferences,
see, e.g., \cite{BBHP.journal}.  

Next, in \cite{RLV08} \bfe{conversion/preference
  games}\index{game!conversion/preference} are introduced.  Such a game
for $n$ players consists of a set $S$ of \bfe{situations} and for
each player $i$ a \bfei{preference relation} $\succeq_i$ on $S$ and a
\bfe{conversion relation} $\myra_i$ on $S$.  The definition is very
general and no conditions are placed on the preference and conversion
relations.  These games are used to formalise gene regulation networks
and some aspects of security.

Finally, let us mention another generalisation of strategic games,
called \bfe{graphical games},\index{game!graphical}
introduced by \cite{KLS01}.  These
games stress the locality in taking a decision.  In a graphical game the
payoff of each player depends only on the strategies of its neighbours
in a given in advance graph structure over the set of players.
Formally, such a game for $n$ players with the corresponding strategy
sets $S_1, \ldots, S_n$ is defined by assuming a neighbour function
\emph{N} that given a player $i$ yields its set of neighbours
$\emph{N}(i)$. The payoff for player $i$ is then a function $p_i$
from $\times_{j \in \emph{N}(i) \cup \{i\}} S_j$ to $\mathbb{R}$.

In all mentioned variants it is straightforward to define the notion
of a Nash equilibrium. For example, in the conversion/preferences
games it is defined as a situation $s$ such that for all players $i$,
if $s \myra_i s'$, then $s' \not\succ_i s$.  However, other introduced
notions can be defined only for some variants.  In particular, Pareto
efficiency cannot be defined for strategic games with parametrised
preferences since it requires a comparison of two arbitrary joint
strategies.  In turn, the notions of dominance cannot be defined for
the conversion/preferences games, since they require the concept of a
strategy for a player.

Various results concerning finite strategic games, for instance the
IESDS Theorem \ref{thm:ies}, carry over directly to the strategic
games as defined in \cite{OR94} or in \cite{ARV08}.  On the other
hand, in the variants of strategic games that rely on the notion of a
preference we cannot consider mixed strategies, since the outcomes of
playing different strategies by a player cannot be aggregated.

\section{Mechanism design}
\label{sec:mech}

Mechanism design is one of the important areas of economics. The 2007
Nobel Prize in Economics went to three economists who laid
its foundations.  To quote from \cite{Eco07},
mechanism design deals with the problem of `how to arrange our
economic interactions so that, when everyone behaves in a
self-interested manner, the result is something we all like'.  So
these interactions are supposed to yield desired social decisions when
each agent is interested in maximising only his own utility.

In mechanism design one is interested in the ways of inducing the players to submit
true information. This subject is closely related to
game theory, though it focuses on other issues. In the next section we shall clarify this connection.
To discuss mechanism design in more detail we need to introduce some basic
concepts.

Assume a set $\{1, \ldots, n\}$ of players
with $n > 1$, a non-empty set of \textbf{\textit{decisions}} $D$, and for each player $i$

\begin{itemize}
\item a non-empty set of \bfe{types}\index{type} $\Theta_i$, and 

\item an \bfe{initial utility function}\index{utility function} 
$
v_i  : D \times \Theta_i \rightarrow \mathbb{R}.
$
\end{itemize}
In this context a type is some private information known only to the player, for example,
in the case of  an auction, the player's valuation of the items for sale. 

When discussing types and sets of types we use then the same abbreviations as in Section \ref{sec:basic}.
In particular, we define
$\Theta := \Theta_1 \times \cdots \times \Theta_n$ and for
$\theta \in \Theta$ we have $(\theta_i, \theta_{-i}) = \theta$.
  



A \textbf{\textit{decision rule}} is a function $f: \Theta \rightarrow D$.
We call the tuple
\[
(D, \Theta_1, \ldots, \Theta_n, v_1, \ldots, v_n, f)
\]
a \textbf{\textit{decision problem}}.

Decision problems are considered in the presence of a \bfe{central authority} who 
takes decisions on the basis of the information provided by the players.
Given a decision problem the desired decision is obtained through the 
following sequence of events, where $f$ is a
given, publicly known, decision rule:

\begin{itemize}

\item each player $i$ receives (becomes aware of) his type $\theta_i \in
  \Theta_i$,

\item each player $i$ announces to the central authority a type 
$\theta'_i  \in \Theta_i$;
this yields a joint type $\theta' := (\theta'_1, \ldots, \theta'_n)$,

\item the central authority then takes the decision $d := f(\theta')$
and communicates it to each player,

\item the resulting initial utility for player $i$ is then
$v_i(d, \theta_i)$.
\end{itemize}

The difficulty in taking decisions through the above described
sequence of events is that players are assumed to be
rational, that is they want to maximise their utility.  As a
result they may submit false information to manipulate the outcome
(decision).  
To better understand the notion of a decision problem consider the following two
natural examples.

\begin{example}\textbf{[Sealed-bid Auction]} \label{exa:auction}
  
  We consider a \bfe{sealed-bid auction}\index{auction!sealed-bid} in which there is a single
  object for sale. Each player (bidder) simultaneously submits to the
  central authority his type (bid) in a sealed envelope and the object
  is allocated to the highest bidder.

Given a sequence $a := (a_1, \ldots, a_j)$ of
reals denote the least $l$ such that $a_l = \max_{k \in \{1,
  \ldots, j\}} a_k$ by $\textrm{argsmax} \: a$.
Then we can model a sealed-bid auction as the following decision problem
$(D,\Theta_1,\ldots,$ $\Theta_n, v_1,\ldots,v_n,f)$:

\begin{itemize}

\item $D = \{1, \ldots, n\}$,

 \item for all $i \in \{1, \ldots, n\}$, $\Theta_i = \mathbb{R}_+$;
$\theta_i \in \Theta_i$ is player's $i$ valuation of the object,

\item for all $i \in \{1, \ldots, n\}$, $v_i(d, \theta_i) := (d = i) \theta$,
where $d = i$ is a Boolean expression with the value 0 or 1,
\item 
$
f(\theta) := \textrm{argsmax} \: \theta.
$
\end{itemize}

Here decision $d \in D$ indicates to which player the object is sold.
Further, $f(\theta) = i$, where 
\[
\mbox{$\theta_i = \max_{j \in \{1, \ldots, n\}}
\theta_j$ and $\forall j \in \{1, \ldots, i-1\} \ \theta_j <
\theta_i$. }
\]
So we assume that in the case of a tie the object is
allocated to the highest bidder with the lowest index.

\HB
\end{example}
\begin{example}\textbf{[Public project problem]} \label{exa:public}
 
  This problem deals with the task of taking a joint decision
  concerning construction of a \bfei{public good},\footnote{In Economics
    public goods are so-called not excludable and non-rival
    goods.  To quote from \cite{Man01}: `People cannot be prevented
    from using a public good, and one person's enjoyment of a public
    good does not reduce another person's enjoyment of it.'} for
  example a bridge.
  Each player reports to the central authority his appreciation of the
  gain from the project when it takes place.  If the sum of the
  appreciations exceeds the cost of the project, the project takes
  place and each player has to pay the same fraction of the cost.
  Otherwise the project is cancelled.

This problem corresponds to the following decision problem,
where $c$, with $c > 0$, is the cost of the project:

\begin{itemize}
 \item $D=\{0,1\}$ (reflecting whether a project is cancelled or takes place),

 \item for all $i \in \{1, \ldots, n\}$, $\Theta_i = \mathbb{R}_+$,

 \item for all $i \in \{1, \ldots, n\}$, $v_i(d,\theta_i):=d(\theta_i-\frac{c}{n})$,
 \item 
$
f(\theta) := \begin{cases}
1 & \textrm{if $\sum_{i=1}^n \theta_i\ge c$}\\
0 & \textrm{otherwise.}
\end{cases}
$
\end{itemize}

If the project
takes place ($d = 1$), $\frac{c}{n}$ is the cost share of the project for each
player. 
\HB
\end{example}

Let us return now to the decision rules.
We call a decision rule $f$ \textbf{\textit{efficient}} if for all $\theta \in
\Theta$ and $d' \in D$
\[
\sum_{i = 1}^{n} v_i(f(\theta), \theta_i) \geq \sum_{i = 1}^{n} v_i(d', \theta_i).
\]
Intuitively, this means that for all $\theta \in \Theta$, $f(\theta)$
is a decision that maximises the \bfe{initial social welfare}\index{social welfare} from a decision 
$d$, defined by 
$
\sum_{i = 1}^{n} v_i(d, \theta_i)
$.
It is easy to check that the decision rules used in Examples \ref{exa:auction} and
\ref{exa:public} are efficient.

Let us return now to the subject of manipulations.  
As an example, consider the case of the
public project problem.  A player whose type (that is, appreciation of
the gain from the project) exceeds the cost share $\frac{c}{n}$
should manipulate the outcome and
announce the type $c$. This will guarantee that the
project will take place, irrespective of the types announced by the
other players.  Analogously, a player whose type is lower than
$\frac{c}{n}$ should submit the type $0$ to minimise the chance that
the project will take place.

To prevent such manipulations we use \bfe{taxes}\index{tax}, which are transfer
payments between the players and central authority.  This leads to a
modification of the initial decision problem $(D, \Theta_1, \ldots,
\Theta_n, v_1, \ldots, v_n, f)$ to the following one:

\begin{itemize}

\item the set of decisions is
$
D \times \mathbb{R}^n,
$

\item the decision rule is a function
$
(f,t): \Theta \myra D \times \mathbb{R}^n,
$
where $
t: \Theta \myra \mathbb{R}^n
$
and
$
(f,t)(\theta) := (f(\theta), t(\theta)),
$

\item the  \bfe{final utility function}\index{utility function}
 of player $i$ is the function $u_i: D \times \mathbb{R}^n \times \Theta_i \myra \mathbb{R}$
defined by
\[
u_i(d,t_1, \ldots, t_n, \theta_i) :=  v_i(d, \theta_i) + t_i.
\] 

\end{itemize}
We call then
$(D \times \mathbb{R}^n, \Theta_1, \ldots, \Theta_n, u_1, \ldots, u_n, (f,t))$ 
a \bfei{direct mechanism} and refer to $t$ as the
\bfei{tax function}.

So when the received (true) type of player $i$ is $\theta_i$ and his
announced type is $\theta'_i$, his final utility is 
\[
u_i((f,t)(\theta'_i,
\theta_{-i}), \theta_i) = v_i(f(\theta'_i, \theta_{-i}), \theta_i) +
t_i(\theta'_i, \theta_{-i}),
\] 
where $\theta_{-i}$ are the types
announced by the other players.

In each direct mechanism, given the vector $\theta$ of announced
types, $t(\theta) := (t_1(\theta), \ldots, t_n(\theta))$ is the vector
of the resulting payments.  If
$t_i(\theta) \geq 0$, player $i$ \emph{receives} from the central authority
$t_i(\theta)$, and if $t_i(\theta) < 0$, he \emph{pays} to the central authority
$|t_i(\theta)|$.

The following definition then captures the idea that taxes prevent manipulations.
We say that a direct mechanism with tax function $t$ is
\bfe{incentive compatible}\index{direct mechanism!incentive compatible} 
if for all $\theta \in \Theta$, $i \in \C{1,\ldots,n}$ and
$\theta'_i \in \Theta_i$
\[
u_i((f,t)(\theta_i, \theta_{-i}), \theta_i) \geq
u_i((f,t)(\theta'_i, \theta_{-i}), \theta_i).
\]
Intuitively, this means that for each player $i$
announcing one's true type ($\theta_i$)
is better than announcing another type ($\theta_i'$).  That is, false
announcements, i.e., manipulations, do not pay off.

From now on we focus on specific incentive compatible direct
mechanisms.  Each \bfei{Groves mechanism} is a direct mechanism
obtained by using a tax function $t(\cdot) := (t_1(\cdot), \ldots, t_n(\cdot))$, where for
all $i \in \C{1, \ldots, n}$

\begin{itemize}

\item $t_i : \Theta \myra \mathbb{R}$ is defined by $t_i(\theta) := g_i(\theta) + h_i(\theta_{-i})$, where

\item $g_i(\theta) := \sum_{j \neq i} v_j(f(\theta), \theta_j)$,
  
\item $h_i: \Theta_{-i} \myra \mathbb{R}$ is an arbitrary function.

\end{itemize}

Note that, not accidentally, 
$v_i(f(\theta), \theta_i) + g_i(\theta)$ is simply the initial social welfare
from the decision $f(\theta)$.

The importance of Groves mechanisms is then revealed by the following crucial result due to \cite{Gro73}.

\begin{theorem}[Groves] \label{thm:groves}
  Consider a decision problem $(D, \Theta_1, \ldots, \Theta_n, v_1,$ $\ldots,
  v_n, f)$ with an efficient decision rule $f$.  Then each Groves
  mechanism is incentive compatible.
\end{theorem}

\Proof
The proof is remarkably straightforward.
Since $f$ is efficient, for all $\theta \in \Theta$, $i \in \{1, \ldots, n\}$ and
$\theta'_i \in \Theta_i$ we have 
\begin{align*}
u_i((f,t)(\theta_i, \theta_{-i}), \theta_i) &=
\sum_{j=1}^{n} v_j(f(\theta_i, \theta_{-i}), \theta_j) +  h_i(\theta_{-i})\\
&\geq \sum_{j=1}^{n} v_j(f(\theta'_i, \theta_{-i}), \theta_j) +  h_i(\theta_{-i}) \\
& = u_i((f,t)(\theta'_i, \theta_{-i}), \theta_i).
\end{align*}
\HB
\VV

When for a given direct mechanism for all $\theta \in \Theta$ we have $\sum_{i = 1}^{n} t_i(\theta) \leq 0$,
the mechanism is called \textbf{\textit{feasible}}, which means that it can be realised without external 
financing.

Each Groves mechanism is uniquely determined by the functions $h_{1}, \ldots$, $h_{n}$.
A special case, called the \bfei{pivotal mechanism}, is obtained by
using
\[
h_i(\theta_{-i}) := - \max_{d \in D} \sum_{j \neq i} v_j(d, \theta_j).
\]

\NI
So then

\[
t_i(\theta)  = \sum_{j \neq i} v_j(f(\theta), \theta_j) - \max_{d \in D} \sum_{j \neq i} v_j(d, \theta_j).
\]

Hence for all $\theta$ and $i \in \{1,\ldots,n\}$ we have
$t_i(\theta) \leq 0$, which means that the pivotal mechanism is feasible
and that each player needs to make the payment $|t_i(\theta)|$ to the
central authority.  

We noted already that the decision rules used
in Examples \ref{exa:auction} and \ref{exa:public} are efficient.
So in each example Groves' Theorem \ref{thm:groves} applies
and in particular the pivotal mechanism is incentive compatible.
Let us see now the details.
\VV

\NI
\textbf{Re: Example \ref{exa:auction}} 
Given a sequence $\theta$ of reals we denote by 
$\theta^*$ its reordering from the largest to the smallest element.
So for example, for $\theta = (1,5,4,3,2)$ we have 
$(\theta_{-2})^{*}_2 = 3$ since $\theta_{-2} = (1,4,3,2)$.

To compute the taxes in the sealed-bid auction in the case of the pivotal mechanism we use the following observation.

\begin{note} \label{not:auction}
In the sealed-bid auction we have for the pivotal mechanism
\[
t_i(\theta) =
 \begin{cases}
  - \theta^{*}_{2}  & \mathrm{if } \ i = \textrm{argsmax} \: \theta \\
       0  & \mathrm{otherwise.}
 \end{cases}
\]
\end{note}
\begin{exercise}
  Provide the proof.
\HB
\end{exercise}
So the highest bidder wins the object and pays for it the amount $\max_{j \neq i} \theta_j$.

The resulting sealed-bid auction was introduced by \cite{Vic61} and
is called a \bfe{Vickrey auction}\index{auction!Vickrey}.
To illustrate it suppose there are three players,
A, B, and C whose true types (bids) are respectively 18, 21, and 24.
When they bid truthfully the object is allocated to player C whose tax (payment) 
according to Note \ref{not:auction} is 21, so the second price offered.
Table \ref{tab:0} summarises the situation.

\begin{table}[htbp]
\normalsize
\begin{center}
\[
\begin{array}{|c|r|r|r|}
\hline
\textrm{player} & \textrm{type} & \textrm{tax} & u_i \\\hline
\hline
A      & $18$ &  $0$    &  $0$\\\hline
B      & $21$ &  $0$    &  $0$\\\hline
C      & $24$ & -$21$ & $3$\\\hline
\end{array}\vspace{0.25cm}
\]
\end{center}
  \caption{The pivotal mechanism for the sealed-bid auction \label{tab:0}}
\end{table}

This explains why this auction is alternatively called
a \bfe{second-price auction}\index{auction!second-price}.  By Groves' Theorem \ref{thm:groves}
this auction is incentive compatible. 
In contrast, the \bfe{first-price auction},\index{auction!first-price} in which the winner pays the
price he offered (so the first, or the highest price), is not
incentive compatible.  Indeed, reconsider the above example. If player C
submits 22 instead of his true type 24, he then wins the object but
needs to pay 22 instead of 24. More formally, in the 
direct mechanism corresponding to the first-price auction we have
\[
u_\textrm{C}((f,t)(18,21,22), 24) = 24 - 22 = 2 > 0 = u_\textrm{C}((f,t)(18,21,24), 24),
\]
which contradicts incentive compatibility for the joint type $(18, 21, 24)$.
\HB

\VV

\NI
\textbf{Re: Example \ref{exa:public}} 
To compute the taxes in the public project problem in the case of the pivotal mechanism we use the following observation.

\begin{note} \label{not:t}
In the public project problem we have for the pivotal mechanism
\[
t_i(\theta) =
 \begin{cases}
       0  & \mathrm{if } \ \sum_{j \neq i} \theta_j \geq  \frac{n-1}{n} c \ \mathrm{ and } \ \sum_{j=1}^{n} \theta_j \geq c \\
       \sum_{j \neq i}\theta_j - \frac{n-1}{n} c  & \mathrm{if } \  \sum_{j \neq i} \theta_j <  \frac{n-1}{n} c \ \mathrm{ and } \ \sum_{j=1}^{n} \theta_j \geq c \\
       0  & \mathrm{if }  \ \sum_{j \neq i} \theta_j \leq  \frac{n-1}{n} c \ \mathrm{ and } \ \sum_{j=1}^{n} \theta_j < c \\
       \frac{n-1}{n} c - \sum_{j \neq i}\theta_j  & \mathrm{if }  \ \sum_{j \neq i} \theta_j > \frac{n-1}{n} c \ \mathrm{ and } \ \sum_{j=1}^{n} \theta_j < c.
 \end{cases}
\]

\end{note}
\begin{exercise}
  Provide the proof.
\HB
\end{exercise}

To illustrate the pivotal mechanism suppose that $c = 30$ and that
there are three players, A, B, and C whose true types are respectively
6, 7, and 25. When these types are announced the project takes place
and Table \ref{tab:1} summarises the taxes that players need to pay
and their final utilities.  The taxes were computed using Note
\ref{not:t}.

\begin{table}[htbp]
\normalsize
\begin{center}
\[
\begin{array}{|c|r|r|r|}
\hline
\textrm{player} & \textrm{type} & \textrm{tax} & u_i \\\hline
\hline
A      & $6$ & $0$    &  -$4$\\\hline
B      & $7$ & $0$    &  -$3$\\\hline
C      & $25$& -$7$ & $8$\\\hline
\end{array}\vspace{0.25cm}
\]
\end{center}
  \caption{The pivotal mechanism for the public project problem \label{tab:1}}
\end{table}

Suppose now that the true types of players are respectively 4, 3 and 22
and, as before, $c = 30$. When these types are also the
announced types, the project does not take place. Still, some players
need to pay a tax, as Table~\ref{tab:2} illustrates.  One can show
that this deficiency is shared by all feasible incentive compatible
direct mechanisms for the public project, see~\cite[ page 861-862]{MWG95}.

\begin{table}[htbp]
\normalsize
\begin{center}
\[
\begin{array}{|c|r|r|r|}
\hline
\textrm{player} & \textrm{type} & \textrm{tax} & u_i \\\hline
\hline
A      & $4$ & -$5$ & -$5$ \\\hline
B      & $3$ & -$6$ & -$6$ \\\hline
C      & $22$ & $0$ & $\phantom{-}0$ \\\hline
\end{array}\vspace{0.25cm}
\]
\end{center}
  \caption{The pivotal mechanism for the public project problem \label{tab:2}}
\end{table}

\section{Pre-Bayesian games}

Mechanism design, as introduced in the previous section, can be
explained in game-theoretic terms using pre-Bayesian games, introduced
by \cite{AMT06} (see also \cite{HB04} and \cite{AB06}).  In strategic
games, after each player selected his strategy, each player knows the
payoff of \emph{every other player}. This is not the case in
pre-Bayesian games in which each player has a private type on which he
can condition his strategy. This distinguishing feature of
pre-Bayesian games explains why they form a class of games with
\emph{incomplete information}.  Formally, they are defined as follows.

Assume a set $\{1, \ldots, n\}$ of players, where $n > 1$. A
\bfe{pre-Bayesian game}\index{game!pre-Bayesian} for $n$ players consists of

\begin{itemize}
\item a non-empty set $A_i$ of \bfe{actions},\index{action}

\item a non-empty set  $\Theta_i$ of \bfe{types},\index{type}

\item a \bfei{payoff function} 
$
p_i : A_1  \times \ldots \times A_n \times \Theta_i \myra \mathbb{R},
$
\end{itemize}
for each player $i$.

Let $A := A_1 \times \ldots \times A_n$.
In a pre-Bayesian game Nature (an external agent) moves first and
provides each player $i$ with a type $\theta_i \in \Theta_i$. Each
player knows only his type.  Subsequently the players simultaneously
select their actions.  The payoff function of each player now depends
on his type, so after all players selected their actions, each player
knows his payoff but does not know the payoffs of the other players.
Note that given a pre-Bayesian game, every joint type $\theta \in
\Theta$ uniquely determines a strategic game, to which we refer below
as a $\theta$-game.

A \bfei{strategy} for player $i$ in a pre-Bayesian game is a
function $s_i : \Theta_i \myra A_i$. 
The previously introduced notions can be naturally adjusted to pre-Bayesian
games. In particular, a joint strategy $s(\cdot) := (s_1(\cdot), \ldots, s_n(\cdot))$
is called an \bfei{ex-post equilibrium} if
\[
\fa \theta \in \Theta \ \fa i \in \{1, \ldots, n\} \ \fa a_i \in A_i \ p_i(s_i(\theta_i), s_{-i}(\theta_{-i}), \theta_i) \geq p_i(a_i, s_{-i}(\theta_{-i}), \theta_i),
\]
where $s_{-i}(\theta_{-i})$ is an abbreviation for the sequence of actions $(s_j(\theta_j))_{j \neq i}$.

In turn, a strategy $s_i(\cdot)$ for
player $i$ is called \bfe{dominant}\index{strategy!dominant} if 
\[
\fa \theta_i \in \Theta_i \ \fa a \in A \ p_i(s_i(\theta_i), a_{-i}, \theta_i) \geq p_i(a_i, a_{-i}, \theta_i).
\]

So $s(\cdot)$ is an ex-post equilibrium iff for every joint type
$\theta \in \Theta$ the sequence of actions
 $(s_1(\theta_1), \ldots, s_n(\theta_n))$ is a Nash equilibrium in the
corresponding $\theta$-game.
Further, $s_i(\cdot)$ is a dominant strategy of player $i$ iff for every type $\theta_i \in \Theta_i$,
$s_i(\theta_i)$ is a dominant strategy of player $i$ in every $(\theta_i, \theta_{-i})$-game.

We also have the following immediate counterpart of
the Dominant Strategy Note \ref{not:dominant}.

\begin{note}[Dominant Strategy] \label{not:dominant1}
Consider a pre-Bayesian game $G$.
Suppose that $s(\cdot)$ is a joint strategy such that each
$s_i(\cdot)$ is a dominant strategy.
Then it is an ex-post equilibrium of $G$.
\end{note}

\begin{example}
As an example of a pre-Bayesian game, suppose that
  \begin{itemize}
  \item $\Theta_1 = \{U, D\}$, $\Theta_2 = \{L, R\}$,

\item $A_1 = A_2 = \{F, B\}$,
  \end{itemize}
and consider the pre-Bayesian game uniquely determined by the following four $\theta$-games. 
Here and below we marked the payoffs in Nash equilibria
in these $\theta$-games in bold.
\pagebreak
\III

\begin{minipage}{10mm}
\VV

  \begin{center}
\VV

  $U$
  \end{center}
\end{minipage}
\begin{minipage}{50mm}
  \begin{center}
\hspace{-14mm}    $L$
  \end{center}
  \begin{game}{2}{2}
         & $F$    & $B$\\
  $F$   & $\textbf{2},\textbf{1}$   &$2,0$\\
  $B$   &$0,1$   &$2,1$
  \end{game}
\end{minipage}
\begin{minipage}{50mm}
  \begin{center}
\hspace{-14mm}    $R$
  \end{center}
  \begin{game}{2}{2}
         & $F$    & $B$\\
  $F$   &$2,0$   &$\textbf{2},\textbf{1}$\\
  $B$   &$0,0$   &$2,1$
  \end{game}
\end{minipage} 
\VV

\begin{minipage}{10mm}
  \begin{center}
\VV

  $D$
  \end{center}
\end{minipage}
\begin{minipage}{50mm}
\begin{game}{2}{2}
       & $F$    & $B$\\
$F$   &$3,1$   &$2,0$\\
$B$   & $\textbf{5},\textbf{1}$   &$4,1$
\end{game}
  \end{minipage}
  \begin{minipage}{50mm}
\begin{game}{2}{2}
       & $F$    & $B$\\
$F$   &$3,0$   &$2,1$\\
$B$   &$5,0$   & $\textbf{4},\textbf{1}$
\end{game}
  \end{minipage}




\III

This shows that the strategies $s_1(\cdot)$ and $s_2(\cdot)$ such that
\[
\mbox{$s_1(\emph{U}) := F, \ s_1(D) := B$, $s_2(L) = F, \ s_2(R) = B$}
\]
form here an ex-post equilibrium.
\HB
\end{example}

However, there is a crucial difference between strategic games
and pre-Bayesian games. 

\begin{example}
Consider the following pre-Bayesian game:

  \begin{itemize}
  \item $\Theta_1 = \{U, B\}$, $\Theta_2 = \{L, R\}$,

\item $A_1 = A_2 = \{C, D\}$.
  \end{itemize}

\III

\begin{minipage}{10mm}
\VV

  \begin{center}
\VV

  $U$
  \end{center}
\end{minipage}
\begin{minipage}{50mm}
  \begin{center}
\hspace{-14mm}    $L$
  \end{center}
\begin{game}{2}{2}
       & $C$    & $D$\\
$C$   &$2,2$   &$0,0$\\
$D$   &$3,0$   &$\textbf{1},\textbf{1}$
\end{game}
\end{minipage}
\begin{minipage}{50mm}
  \begin{center}
\hspace{-14mm}    $R$
  \end{center}
\begin{game}{2}{2}
       & $C$    & $D$\\
$C$   &$2,1$   &$0,0$\\
$D$   &$3,0$   &$\textbf{1},\textbf{2}$
\end{game}
\end{minipage} 
\VV

\begin{minipage}{10mm}
  \begin{center}
\VV

  $B$
  \end{center}
\end{minipage}
\begin{minipage}{50mm}
\begin{game}{2}{2}
       & $C$    & $D$\\
$C$   &$\textbf{1},\textbf{2}$   &$3,0$\\
$D$   &$0,0$   & $2,1$
\end{game}
  \end{minipage}
  \begin{minipage}{50mm}
\begin{game}{2}{2}
       & $C$    & $D$\\
$C$   &$\textbf{1},\textbf{1}$   &$3,0$\\
$D$   & $0,0$ & $2,2$
\end{game}
  \end{minipage}





\III
  
Even though each $\theta$-game has a Nash equilibrium, they are so
`positioned' that the pre-Bayesian game has no ex-post equilibrium.
Even more, if we consider a mixed extension of this game, then the
situation does not change. The reason is that no new Nash equilibria
are then added to the `constituent' $\theta$-games. (Indeed, each of
them is solved by IESDS and hence by the IESDMS Theorem
\ref{thm:imes}$(ii)$ has a unique Nash equilibrium.)  This shows that
a mixed extension of a finite pre-Bayesian game does not need to have
an ex-post equilibrium, which contrasts with the existence of Nash
equilibria in mixed extensions of finite strategic games.  
\HB
\end{example}

To relate pre-Bayesian games to mechanism design we need one more
notion. We say that a pre-Bayesian game is of a
\bfe{revelation-type}\index{game!revelation-type pre-Bayesian} if
$A_i = \Theta_i$ for all $i \in \C{1,\ldots,n}$.  So in a
revelation-type pre-Bayesian game the strategies of a player are the
functions on his set of types.  A strategy for player $i$ is called
then \bfe{truth-telling}\index{strategy!truth-telling} if it is the identity function
$\pi_i(\cdot)$ on $\Theta_i$.

Now, as explained in \cite{AMT06} mechanism design can be viewed as an
instance of the revelation-type pre-Bayesian games. Indeed, we have the following
immediate, yet revealing observation.

\begin{theorem}
Given a direct mechanism   
\[
(D \times \mathbb{R}^n, \Theta_1, \ldots, \Theta_n, u_1, \ldots, u_n, (f,t))
\]
associate with it a revelation-type pre-Bayesian game,
in which each payoff function $p_i$ is defined by 
\[
p_i((\theta_i',\theta_{-i}), \theta_i) := u_i((f, t)(\theta_i',\theta_{-i}), \theta_i). 
\]
Then the mechanism is incentive compatible
iff in the associated pre-Bayes\-ian game for each player truth-telling
is a dominant strategy.
\end{theorem}

By Groves's Theorem \ref{thm:groves} we conclude that in the pre-Bayesian game
associated with a Groves mechanism, $(\pi_1(\cdot), \ldots,
\pi_n(\cdot))$ is a dominant strategy ex-post equilibrium.

\section{Conclusions}

\subsection{Bibliographic remarks}

Historically, the notion of an equilibrium in a strategic game
occurred first in \cite{Cou38} in his study of production levels of a
homogeneous product in a duopoly competition.  The celebrated von
Neumann's Minimax Theorem proved by \cite{Neu28} establishes an
existence of a Nash equilibrium in mixed strategies in two-player
zero-sum games.  An alternative proof of Nash's Theorem, given in
\cite{Nas51}, uses Brouwer's Fixed Point Theorem.

Ever since Nash established his celebrated theorem, a search has
continued to generalise his result to a larger class of games. A
motivation for this endeavour has been the existence of natural infinite
games that are not mixed extensions of finite games.  As an example of
such an early result let us mention the following theorem due to
\cite{Deb52}, \cite{Fan52} and \cite{Gli52}.

\begin{theorem}
Consider a strategic game such that 

\begin{itemize}
\item each strategy set is a non-empty compact convex subset of a
  complete metric space,

\item each payoff function $p_i$ is continuous and quasi-concave in
  the $i$th argument.\footnote{Recall that the function $p_i: S \myra \mathbb{R}$ is
    \bfe{quasi-concave in the $i$th argument}\index{quasi-concave function}
if the set
    $\C{s'_i \in S_i \mid p_i(s'_i, s_{-i}) \geq p_i(s)}$ is convex
    for all $s \in S$.}
\end{itemize}
Then a Nash equilibrium exists.
\end{theorem}

More recent work in this area focused on the existence of Nash equilibria
in games with non-continuous payoff functions, see in particular
\cite{Ren99} and \cite{Bic06}.  

The issue of complexity of finding a Nash equilibrium has been a long
standing open problem, clarified only recently, see \cite{DGP09} for
an account of these developments.
Iterated elimination of strictly dominated strategies and of weakly
dominated strategies was introduced by \cite{Gal53} and \cite{LR57}.
The corresponding results summarised in Theorems \ref{thm:ies}, 
\ref{thm:iter2}, \ref{thm:imes} and \ref{thm:imewds} are folklore results.

\cite{Apt04} provides uniform proofs of various order independence
results, including the Order Independence Theorems \ref{thm:strict}
and \ref{thm:strictmixed}.  The computational complexity of iterated
elimination of strategies has been studied starting with \cite{KPT88},
and with \cite{BBFH09} as a recent contribution.


There is a lot of work on formal aspects of common knowledge and of its consequences for game theory.
see, e.g., \cite{Aum99} and \cite{BB99}.

\subsection{Suggestions for further reading}

Strategic games form a large research area and we have barely scratched its surface.
There are several other equilibria notions and various other types of games.

Many books provide introductions to various areas of game theory,
including strategic games.  Most of them are written from the
perspective of applications to Economics.  In the 1990s the leading
textbooks were \cite{Mye91}, \cite{Bin91}, \cite{FT91} and
\cite{OR94}.

Moving to the next decade, \cite{Osb05} is an excellent, broad in its
scope, undergraduate level textbook, while \cite{Pet08} is probably
the best book on the market on the graduate level.  Undeservedly less
known is the short and lucid \cite{Tij03}.  An elementary, short
introduction, focusing on the concepts, is \cite{SLB08}.  In turn,
\cite{Rit02} is a comprehensive book on strategic games that also
extensively discusses \bfe{extensive games}, i.e., games in which the
players choose actions in turn.  Finally, \cite{Bin07} is a thoroughly
revised version of \cite{Bin91}.

Several textbooks on microeconomics include introductory chapters on
game theory, including strategic games.  Two good examples are
\cite{MWG95} and \cite{JR00}.  Finally, \cite{NRTV07} is a recent
collection of surveys and introductions to the computational aspects
of game theory, with a number of articles concerned with strategic
games and mechanism design.

 \bibliography{/ufs/apt/bib/e,/ufs/apt/bib/rossi05}

\bibliographystyle{abbrvnat}

\end{document}

%% file: mydefinitions.tex
\usepackage[english]{babel}
\usepackage{graphicx}
\usepackage{amsthm}
\usepackage{proof}
\usepackage{amsmath}
\usepackage{amssymb}
\usepackage[dvips]{color}
\usepackage{enumerate}
\usepackage{url}

\newcommand{\Proof}{\NI
                    {\bf Proof.}\ }

\newcounter{symbol}
\newcommand{\indexsyma}[1]%
{\stepcounter{symbol}\index{zzz1 \thesymbol @\protect#1}}
\newcommand{\indexsymb}[1]%
{\stepcounter{symbol}\index{zzz2 \thesymbol @\protect#1}}
\newcommand{\indexsymc}[1]%
{\stepcounter{symbol}\index{zzz3 \thesymbol @\protect#1}}
\newcommand{\indexsymd}[1]%
{\stepcounter{symbol}\index{zzz4 \thesymbol @\protect#1}}
\newcommand{\indexsyme}[1]%
{\stepcounter{symbol}\index{zzz5 \thesymbol @\protect#1}}

\newcommand{\bfe}[1]{\begin{bfseries}\emph{#1}\end{bfseries}\index{#1}}

\newcommand{\bfei}[1]{\begin{bfseries}\emph{#1}\end{bfseries}\index{#1}}

\newcommand{\myra}{\mbox{$\:\rightarrow\:$}}

\newcommand{\sse}{\mbox{$\:\subseteq\:$}}

\newcommand{\fa}{\mbox{$\forall$}}
\newcommand{\te}{\mbox{$\exists$}}

\newcommand{\C}[1]{\mbox{$\{{#1}\}$}}           

\newcommand{\NI}{\noindent}
\newcommand{\HB}{\hfill{$\Box$}}
\newcommand{\VV}{\vspace{5 mm}}
\newcommand{\III}{\vspace{3 mm}}
\newcommand{\II}{\vspace{2 mm}}

\newcommand{\szkew}[1]{\relax \setbox0=\hbox{\kern -24pt $\displaystyle#1$\kern 0pt }%
\box0}
{\catcode`\@=11 \global\let\ifjusthvtest@=\iffalse}

\newcounter{oldmycaption}



%% file: line.pstex_t
\begin{picture}(0,0)%
\includegraphics{line.pstex}%
\end{picture}%
\setlength{\unitlength}{3947sp}%
\begingroup\makeatletter\ifx\SetFigFont\undefined%
\gdef\SetFigFont#1#2#3#4#5{%
  \reset@font\fontsize{#1}{#2pt}%
  \fontfamily{#3}\fontseries{#4}\fontshape{#5}%
  \selectfont}%
\fi\endgroup%
\begin{picture}(3024,174)(2539,-1648)
\end{picture}%

%% file: line2.pstex_t
\begin{picture}(0,0)%
\includegraphics{line2.pstex}%
\end{picture}%
\setlength{\unitlength}{3947sp}%
\begingroup\makeatletter\ifx\SetFigFont\undefined%
\gdef\SetFigFont#1#2#3#4#5{%
  \reset@font\fontsize{#1}{#2pt}%
  \fontfamily{#3}\fontseries{#4}\fontshape{#5}%
  \selectfont}%
\fi\endgroup%
\begin{picture}(3024,924)(2539,-2023)
\end{picture}%